\documentclass[12pt,a4]{article}
\usepackage[utf8]{inputenc}
\usepackage[T1]{fontenc}
\usepackage{amsmath, amssymb, dsfont,amsthm}
\usepackage[normalem]{ulem}  
\usepackage{enumitem}
\usepackage{graphicx}
\usepackage{mathrsfs}%
\usepackage{xcolor}%
\usepackage{booktabs}%
\usepackage{algorithm}%
\usepackage{algorithmicx}%
\usepackage{algpseudocode}%

\usepackage{listings}%

\newcommand{\Dcal}{\mathcal{D}}

\newcommand{\Lcal}{\mathcal{L}}

\newcommand{\Xcal}{\mathcal{X}}

\newcommand{\R}{\mathbb{R}}
\newcommand{\E}{\mathbb{E}}

\newcommand{\util}{\tilde{u}}

\newcommand{\Ttil}{\tilde{T}}
\newcommand{\tinit}{\tilde{T}_{init}}
\newcommand{\csigmatil}{\tilde{c}_\sigma}
\newcommand{\csbtil}{\tilde{c}_{sb}}
\newcommand{\cvtil}{\tilde{c}_v}

\newcommand{\dloss}{D_{\mathrm{logMSE}}}

\newtheorem{remark}{Remark}%

\title{Physics-Informed Neural Networks for coupled stiff transport systems}

\author{
Laetitia Laguzet\thanks{CEA-DAM-DIF, F-91297 Arpajon, France, \text{Latitia.Laguzet@cea.fr},ORCID  0009-0000-9655-0880} 
\and
Gabriel Turinici\thanks{CEREMADE, Université Paris Dauphine - PSL,  Paris 75116, Paris, France; \text{Gabriel.Turinici@dauphine.fr}, \text{https://turinici.com}, ORCID 0000-0003-2713-006X}
}
	
\date{Oct. 14th, 2026}

\begin{document}
\maketitle
	
\begin{abstract}
%
%
%
Physics-Informed Neural Networks (PINN) struggle with stiff, re\-gime-changing transport equations due to instability, loss imbalance, and violations of physical consistency. This paper investigates these failures through the Marshak wave equations, a canonical benchmark from radiative transport, where initial and boundary conditions differ by up to 12 orders of magnitude, and proposes modifications to the standard PINN framework to overcome them.
Three modifications are introduced: a ScaledSigmoid final activation enforcing physical bounds and positivity of the unknowns; a logarithmic MSE loss replacing the standard quadratic loss for initial and boundary conditions, enabling training across extreme scale disparities; and explicit enforcement of global conservation laws as an additional physics loss term. Monte Carlo sampling with exponential time weighting is used throughout.
The proposed framework recovers the Marshak wave dynamics - including the hot, cold, and wave-front regions - in good agreement with a reference Implicit Monte Carlo solution (sub-millimeter absolute error on the front position, although the relative error remains substantial at early times), with run times under 5 minutes. Ablation studies confirm that each ingredient is essential: linear activation, absence of the logarithmic loss, or removal of the PDE term each independently cause the method to fail qualitatively.
This work identifies and resolves three concrete failure modes of standard PINN on stiff hyperbolic systems with nonlinear coupling. The combination of bounded activations, scale-aware loss functions, and conservation law enforcement constitutes a framework validated on the present one-dimensional Marshak wave problem, with potential extension to other coupled stiff transport systems in engineering.

\end{abstract}
	
\section{Motivation}

Physics-Informed Neural Networks (PINN) are a class of neural networks designed to integrate physical laws directly into the learning process 
 allowing to learn solutions that automatically respect those underlying equations.
By embedding prior knowledge in the form of governing equations, PINN enable the modeling and prediction of complex physical systems even when available data are limited or incomplete. For such systems, the governing equations (typically ordinary or partial differential equations), together with initial and boundary conditions, are enforced as soft constraints in the training objective.

PINN were introduced by Raissi et al.~\cite{raissi2019pinns} and have since found applications in physics, engineering, and fluid mechanics, in particular for solving partial differential equations (PDEs) and other dynamical systems \cite{karniadakis2021physics}.

In fluid dynamics, PINN have been applied to canonical and complex flows governed by the Navier--Stokes equations, including laminar and transitional or moderately turbulent regimes, as well as advection--diffusion--reaction systems \cite{sun2020surrogate,jagtap2020adaptive}. Their ability to incorporate boundary conditions and other physical constraints
is especially valuable in regimes where experimental or high-fidelity simulation data are scarce.

Beyond forward simulations, PINN are particularly well suited for inverse problems, where unknown physical parameters, source terms, or constitutive relations must be identified from partial and noisy observations while preserving physical consistency \cite{raissi2018hidden}. This capability makes them attractive for data assimilation, system identification, and optimization tasks in multi-physics settings.

PINN can also serve as physics-consistent surrogate models for uncertainty quantification and rapid evaluation of parameter-dependent PDEs, offering advantages over purely data-driven surrogates, particularly in extrapolative or data-scarce regimes \cite{yang2021bpinn,inversepinn_berardi25}.

Recent research increasingly focuses on improving the robustness and scalability of PINN for multiscale and stiff problems, for which standard formulations often fail \cite{wang2023when,baty2023solvingstiffordinarydifferential,bento2026solvingstiffdarkmatter,hofmann2026mitigatingnumericalstiffnessleastsquares,jesser2024numericalrobustnesspinnsmultiscale}. Active research directions include adaptive loss balancing strategies, domain decomposition approaches, physics-informed operator learning frameworks such as DeepONets and neural operators, and hybrid methods combining classical numerical solvers with PINN components \cite{lu2021deeponet,li2021fourier}. Parallel efforts in Bayesian PINN and uncertainty quantification aim to provide calibrated predictions and stronger theoretical guarantees, reinforcing the role of PINN as reliable tools in scientific computing \cite{yang2021bpinn,karniadakis2021physics}.

\subsection{Our case and short literature review}

Within this ongoing effort we focus in this contribution on stiff transport equations that proved to be challenging for the usual implementations of PINN.
There are not many use cases of PINN for transport equations similar to the ones we are interested in stiff, regime changing evolution; 
 among those, 
Wang et al. \cite{WANG2022109234} 
discuss a case of neutron transport (and also eigenvalue-type cases) and introduce the "conservative PINN" (cPINN), working in particular on interface conditions between subdomains in a specific way; they mention that imposing boundary conditions is very delicate (see Section 3.3); however, the constants make their regime still different and less dynamic than ours (see next section).

On the other hand Liang et al. \cite{liang_continuous_2024} 
 focus on conservative-form equations, whether time-dependent or not, and note that PINN have difficulties finding good solutions; they explicitly mention the case of quotient-type equations (such as ours, where one unknown quantity divides another) where they state they cannot provide a solution because it is physically inconsistent.

Motivated by the same empirical remarks, \cite{pinn_gradient_pathologies21}, 
written by one of the inventors of the method, observe that the terms of the loss functional $\mathcal{L}(\cdot)$ in 
\eqref{eq:fonctionlosspinn1}-\eqref{eq:fonctionlosspinn2}
are very fragile to optimize together and that care must be taken in choosing the best coefficients $c_\phi$, $c_{ic}$, $c_{bc}$ to use in 
\eqref{eq:fonctionlosspinn1}-\eqref{eq:fonctionlosspinn2}.

In this general context we propose adaptations to the PINN framework that we find necessary to solve the stiff, regime-changing transport equations of the present investigation. With respect to the existing literature we introduce specific activation functions, a logarithmic MSE loss to treat the initial and boundary conditions, and explicitly include conservation laws in the loss functional.

	The balance of the paper is as follows: we present in Section
\ref{sec:pinn_notations} the main notations for PINN and in Section~\ref{sec:eq_Marshak} the Marshak equations. The methodology including the original modifications to PINN framework
is presented in Section
\ref{sec:methodology}; the numerical results are the object of Section~\ref{sec:results} followed by concluding remarks in Section~\ref{sec:conclusion}.

\section{PINN notations} \label{sec:pinn_notations}

PINN (Physics-Informed Neural Networks) rely on the use of a multi-layer neural network to approximate a function $u(t,x)$ that satisfies the following constraints:
\begin{itemize}
	\item it solves some ordinary differential equation (ODE) or partial differential equation (PDE),
	\item has associated initial conditions and boundary conditions,
	\item is coherent with (possibly partial) measurements within the domain or other observations,
	\item certain specific constraints like positivity, conserved quantities and so on.
\end{itemize}

\begin{figure}
	\begin{center}
		\includegraphics[height=4cm]{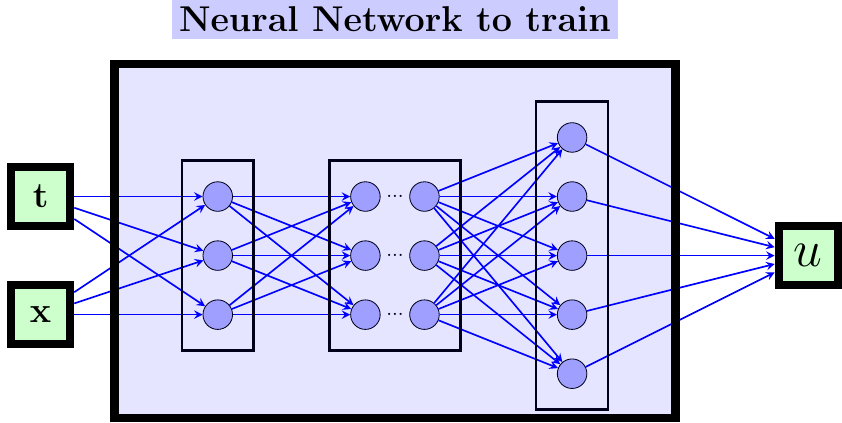}
	\end{center}
	\caption{\small \it Illustration of a neural network generating the solution as a function (to be learned) of the inputs. The advantage is that standard ``deep learning'' libraries can easily compute the derivatives of the output with respect to the inputs and also with respect to the network parameters.}
	\label{fig:pinn_nn}
\end{figure}

Let $\theta$ be the set of parameters of the neural network that 
will construct the approximation of the function $u$; we designate  $u_\theta$ this approximation; more precisely, the NN takes as input  a generic couple $(t,x)$  and outputs a real number (or vector) $u_\theta(t,x)$  as in Figure~\ref{fig:pinn_nn}, that we hope close to the real solution $u(t,x)$.

Let a general PDE be given in the form:
\begin{align}
	\mathcal{F}[u](t,x) = 0, \quad x \in \Xcal, \quad t \in [0, T],
\end{align}
where $\mathcal{F}$ is the differential operator describing the physics of the problem and $\Xcal$ the spatial domain, PINN aim to minimize a loss function defined as follows:
\begin{align}
	\mathcal{L} = \mathcal{L}_\text{data} + \mathcal{L}_\text{physics},
\end{align}
where:
\begin{itemize}
	\item $\mathcal{L}_\text{data}$ represents the error with respect to observed data: initial conditions, boundary conditions, other measurements or observations; 
	the discrepancy between the network predictions and the training data is measured, for example, via the mean squared error; this is where boundary or initial conditions are taken into account (see \cite{hard_constraintsPINN21} for a more structural approach to these conditions).
	\item $\mathcal{L}_\text{physics}$ enforces that the solution $u(t, x)$ satisfies the physical laws described by $\mathcal{F}u$.
	For instance, assume we look for the  solution of $\partial_t u + a \partial_x u =0$; with previous notations:
	\begin{align}
		\mathcal{F}[u](t,x)=\partial_t u(t,x) + a \partial_x u(t,x).
	\end{align} 
and the equation error is 
	\begin{align}
		f_\theta(t,x):=\mathcal{F}[u_\theta](t,x)=\partial_t u_\theta(t,x) + a \partial_x u_\theta(t,x).
	\end{align}
	In the continuous version, the term $\mathcal{L}_\text{physics}$ could contain:
	\begin{align}
\mathcal{L}_\text{physics}=		\int_\Xcal \int_0^T \|f_\theta(t,x)\|^2 \, dt \, dx \ + \text{ (other similar terms)}
	\end{align}
	whereas in the discrete version, it contains, for example, the error of the differential equation evaluated at collocation points $t_i,x_i$:
	\begin{align}
		\mathcal{L}_\text{physics}^\text{discrete} = \frac{1}{N} \sum_{i=1}^N \|f_\theta(t_i,x_i)\|^2 \ + \text{ (other similar terms)},
\label{eq:l_physics_discrete}
	\end{align}
\end{itemize}
The term 
$\mathcal{L}_\text{physics}^\text{discrete} $
may also contain some other terms
 that describe important physical laws of the evolution system. Putting together all the parts, the PINN solution is the minimizer of the following function (we disregard smoothness conditions for now)~:
\begin{align}
	&\mathcal{L}(\theta) :=	
	c_{\phi}\underbrace{\int_\Xcal \int_0^T f_\theta(t,x)^2 dt dx +\text{ (other similar terms)}
	}_{\mathcal{L}_\text{physics}} 
	\label{eq:fonctionlosspinn1}
	\\ &
	+ \underbrace{ 
		c_{ic} \cdot
		\Lcal_{ic}(u_\theta(0,x) , \xi(x))
+ 
		c_{bc}	\cdot 	\Lcal_{bc}(u_\theta(t,x) , u_{bc}(t,x))
	}_{\mathcal{L}_\text{data}}, 
	\label{eq:fonctionlosspinn2}
\end{align}
where $c_\phi$, $c_{ic}$ et $c_{bc}$  are positive constants that translate relative optimization priorities of the three parts of the loss functional $\Lcal(\cdot)$.

The loss functions in \eqref{eq:fonctionlosspinn2} convey the requirement that 
$u_\theta(0,x)$ needs to be close to $\xi(x)$ (initial condition) and 
$u_\theta(t,x)$ to $u_{bc}(t,x)$ (boundary conditions); popular candidates are the mean square integrals:
\begin{align}
&\Lcal_{ic}^{L^2}(u_\theta(0,x) , \xi(x))= \int_\Dcal (u_\theta(0,x) - \xi(x) )^2 dx \\
&	\Lcal_{bc}^{L^2}(u_\theta(t,x) , u_{bc}(t,x))=\int_{[0,T] \times\partial\Dcal} (u_\theta(t,x) - u_{bc}(t,x) )^2 dxdt,
\label{eq:loss_ic_bc}
\end{align}
but other choices can be made.  

The original reference \cite{raissi2019pinns} is highly cited in the numerical modeling and computational physics literature. Several extensions have been proposed to address certain limitations of the initial formulation. For example, among the most recent ones, PIKANs propose an extension beyond classical neural networks, see the recent article \cite{toscano2024pinnspikansrecentadvances}. Here we present a first approach for the case of coupling a transport equation with another physical measurement.

\subsection{Neural network component of PINN}

PINN rely on neural networks to approximate the solution $u(t,x)$ of the differential equation. The neural network is composed of multiple layers, using suitable nonlinear activation functions (for example, ``ReLU3'' or ``$\tanh$'') to capture complex structures in the data. The role of the network is twofold: to approximate the unknown solution $u(t,x)$ in terms of the input variables (for example, spatial coordinates $x$ and temporal coordinates $t$) and to directly integrate the physical laws into the optimization process via the loss function.

The PINN approach leverages the fact that neural networks are implemented in libraries such as TensorFlow (Keras), PyTorch, etc., which provide automatic differentiation, i.e., the ability to automatically compute the derivative of any function coded within the library. Moreover, it has been shown that neural networks are sufficiently expressive to approximate any desired function.

\section{Stiff hyperbolic systems: the Marshak wave example}
\label{sec:eq_Marshak}

Although PINN were seen to obtain good results in general, the situation of non-linear coupling of PDEs has not been previously treated, especially for stiff hyperbolic systems. Interested by a Marshak wave application, we will consider in this work the following PDE system, see~\cite{	marshak1958effect,mcclarren2008effects,LAGUZET_Turinici_24_quantization,marshak22_imex,marshak2B_2015,marshak2B_2022,xingMonteCarlo25}:

\begin{align}
& \forall \ t\le T, x\in \Xcal, \omega \in \Omega : 
	\nonumber \\
	& \frac{1}{c} \partial_t \util(t,x,\omega)+ \omega 
	\nabla_x \util(t,x,\omega) + \frac{\csigmatil}{\Ttil^3(t,x)} \util(t,x,\omega) =  \csbtil \Ttil(t,x)  
	\label{eq:tilde_eq_uT1}
	\\ &
	\cvtil \partial_t \Ttil(t,x) =  \frac{\csigmatil}{\Ttil^3(t,x)} \langle\util\rangle(t,x)-\csbtil \Ttil(t,x), \  \forall t\le T, x\in \Xcal
	\label{eq:uTtilde} 
	\\ &
	\Ttil (t=0,x)= \tilde{T}_{init}, \forall x \in \Xcal
\label{eq:T_init_conditions}
\\ &  
	\Ttil (t,x=x_{min})=\tilde{T}_{leftbct}, \ \forall t \le T
\label{eq:T_left_conditions}
\\ & 
	\util(t=0,x,\omega)= \tilde{u}_{init}, \ \forall x\in \Xcal, \omega \in \Omega
\label{eq:u_init_conditions}
\\ & 
	\util(t,x_{max},\omega)= \tilde{u}_{init}, \ \forall t \le T, \omega \in \Omega
\label{eq:u_xmax_conditions}
\\ & 
	\util(t,x_{min},\omega)=\tilde{u}_{leftbcu}, \forall \omega \in \Omega_x,
	\label{eq:tilde_eq_uT_last}
\end{align}
with the constants in the Table~\ref{tab:valeursMarshak}, 
and where
\begin{itemize}
\item 
 for any $\psi$:
\begin{align}
	\langle{\psi}(t,x) \rangle:= \E_{\omega \text{ uniform over } \Omega }[\psi(t,x,\omega)].
	\label{eq:mean_value_psi}
\end{align}
\item
the spatial domain is  $\Xcal=[x_{min},x_{max}]$
\item  the time domain is $t\in [0,T]$ 
\item 
$\omega \in  \Omega$ designates propagation directions; using the $S_N$ formalism (see~\cite{carlson_solution_1958}) we will take $\Omega$ to be~:
\begin{equation}
	\Omega_N= \{\omega_k=\cos(2 \pi k/N);k=0,...,N-1\}.
	\label{eq:def_Omega}
\end{equation}
When the number of directions is implicitly known we will just write $\Omega$ instead of $\Omega_N$.
\item 
$\partial \Xcal=\{ x_{min},x_{max}\}$ is the boundary of $\Xcal$ 
\item $ \partial \Omega_x$ is the set of directions that enter $\Xcal$ at the point  $x$ of its border:
\begin{equation}
	\Omega_x = \{ \omega \in \Omega: \langle \omega, x \rangle \ge 0 \}.
\end{equation}
\item the functions  $\tilde{u}(x,t,\omega)$
 and $\tilde{T}(x,t) \in \mathbb{R}$ are the unknowns; often 
we write $\tilde{u}(x,t) \in \mathbb{R}^N$ to designate the vector 
$(\tilde{u}(x,t,\omega))_{\omega \in \Omega_N}$;
\item the 
Equations \eqref{eq:T_init_conditions}-\eqref{eq:tilde_eq_uT_last}
specify the initial and boundary conditions; the dynamics is generated by entering particles through the left boundary; we suppose the right boundary is far enough such that the wave front does not reach it before $T$.
\end{itemize}

\noindent One of the difficulties of this setting lies in the  difference of $12$ orders of magnitude  between the boundary conditions
	$\tilde{u}_{leftbcu} $
 and the initial conditions 
 	$\tilde{u}_{init} $
 for  $\util$.

\begin{table*}[!htb]
	\centering
	\begin{tabular}{|c|c|c|c|c|}
		\hline
		$x_{min}$ & $0$(cm) & $x_{max}$&$0.5$ (cm)
		\\ \hline 
		$T$ & $10$ (ns) & N & 8 \\ \hline
		c & 30 (cm/ns)  &$\csigmatil$ &468.75 ($cm^{-1} \cdot MK^3$)
		\\ \hline 
		$\csbtil $ & 531.5625 ($g \cdot cm^{-1} \cdot ns^{-3}$) &
		$\cvtil$ & 1.292655 
		$ \left(\frac{g}{cm\cdot ns^{2} \cdot MK}\right)  $
		\\ \hline 
		$\tilde{T}_{init}$ & $1.1604 \cdot 10^{-3}$ (MK) &
		$\tilde{T}_{leftbct}$ & $1.1604$ (MK) 
		\\ \hline 
		$\tilde{u}_{init} $ & $2.0561 \cdot 10^{-12}$ &
		$\tilde{u}_{leftbcu} $ & $2.0561$
		\\ \hline 
$n_{neurons}$ & 32 & $n_{batch}$ & 1024 $\cdot$ 8
		\\ \hline 
$n_{cons}$ & 10 & $n_{MonteCarlo}$ & 1000
		\\ \hline 
$	v_{min}$ for $\tilde{u}$ & $0.999 \cdot \tilde{u}_{init} \text{ (fixed) }$
& $v_{max}$ for $\tilde{u}$ &$ 1.001 \cdot \tilde{u}_{leftbcu}
\text{ (fixed) }$
	\\ \hline 
$	v_{min}$ for $\tilde{T}$ & $0.999 \cdot \tilde{T}_{init} \text{ (fixed) }$
& $v_{max}$ for $\tilde{T}$ &$ 1.001 \cdot \tilde{T}_{leftbct}
\text{ (fixed) }$
	\\ \hline 
$c_{cons}$ & $10^6$ & $c_\phi$& $10^2$ for $\util$, $10^2$ for $\Ttil$ 
		\\ \hline 
$c_{ic}$ & $10^5$ & $c_{bc}$& $10^5$ 
\\ \hline 
$\lambda$ & $1/T$ & $\mathcal{T}$ & $\{ 1ns,5ns,10ns \}$ 
\\ \hline 
$n_{iter}$ & $10{,}000$ & optimizer& Adam 
\\ \hline 
$\epsilon_L$ & $10^{-14}$ & &  
\\ \hline 
	\end{tabular}
	\caption{\it \small Values and units used in the numerical simulation of the propagation of a Marshak-type wave in an opaque medium, reference case. \label{tab:valeursMarshak}}
\end{table*}

\begin{remark}
Note that boundary and initial conditions for $\tilde{u}$ are incoherent, i.e., there are discontinuous at $t=0, x=x_{min}$ where the $t=0$ side is of order $10^{-12}$ while the $x=x_{min}$ side is of order $1$. This loss of continuity is expected for such stiff transport equations and will make the task of the NN difficult because it has to fit both data.
\end{remark}	

\begin{remark}
With respect to standard Marshak formulas~\cite{
	marshak1958effect,mcclarren2008effects,
	LAGUZET_Turinici_24_quantization} we operated as rescaling on the usual intensity and temperature variables.
To obtain the standard variables denoted $u$, and $T$, use the relations:
\begin{equation}
	\boxed{u = \util \cdot 10^{-3}}, \ \ \ \
	\boxed{T=10\cdot \tilde{T} }.
\end{equation}
\end{remark}	
\begin{remark}
The boundary condition \eqref{eq:T_left_conditions} on $\Ttil$ is not a well-posedness requirement of the hyperbolic system: the temperature at the entrance is determined by the coupled dynamics of the system. It encodes, however, a physical fact ( the medium at the left boundary relaxes toward the equilibrium temperature of the incoming radiation) and it provides the network with additional supervisory information to be reconciled with the PDE residual and the other data. We impose it in the loss function: it does not over-determine the system and, in practice, helps the training; the results of Section~\ref{sec:sensitivity} indicate that the solution is stable with respect to this choice.
\end{remark}

\section{Methodology} \label{sec:methodology}

Let us now describe to the PINN implementation in detail and the insights into the construction of an efficient NN architecture.

\subsection{Network architecture} \label{sec:network_arch}

\begin{figure}
	\begin{center}
		\includegraphics[width=\textwidth]{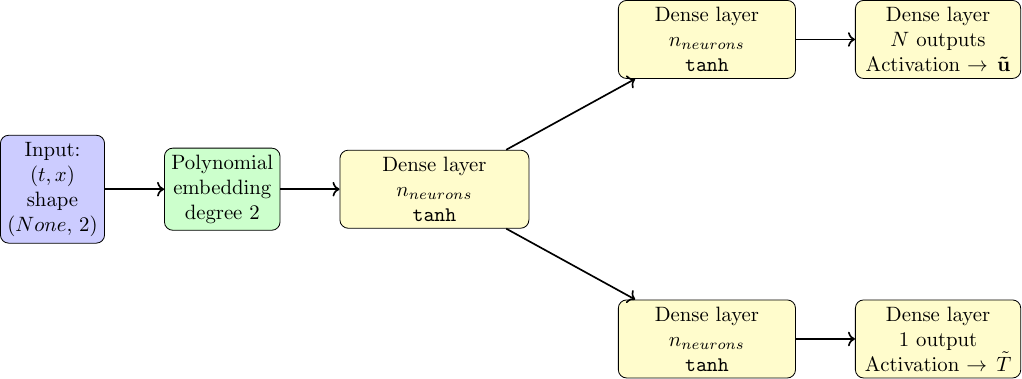}
	\end{center}
	\caption{\small \it Actual architecture of the  neural network.}
	\label{fig:pinn_nn_branched}
\end{figure}

First note that here we have two unknowns, $\tilde{u}(t,x)\in \R^N $ and $\tilde{T}(t,x) \in \R$; a first architecture decision is whether one should create a single network that outputs  $\tilde{u}$ jointly with $\tilde{T}$. In this case the input size would be $2$ (one dimension for $t$ and other for $x$) and the output $N+1$. The alternative would be to have two disjoint networks, one for $u$ and the other for $T$.  

The advantage of having one network is simplicity but also that $\util$ and $\Ttil$ share some common features learned in the initial layers of the network.  After some preliminary numerical tests (not shown here) 
we concluded that, even if the possibility of having two distinct networks
may enhance the results for some parameter values, the critical advancements come from other choices so we retain the option of having a common trunk network followed for both $u$ and $T$ by a specific part.
As it is classic in PINN we 
also added an embedding, here a polynomial embedding of order $2$ which means that the input $t,x$ is transformed into a vector $t,x,t^2, tx, x^2$; combined with a fully connected architecture this gives the layout (see Figure~\ref{fig:pinn_nn_branched}):
\begin{enumerate}
	\item 
	Network structure: input of size (None, 2), ('None' is a placeholder for the batch size) followed by a polynomial of degree 2 embedding and a dense layer with $n_{neurons}$, `tanh' activated; the output of this layer is then feed to 
	two different networks:
	\begin{enumerate}
		\item 	 a linear, tanh activated layer with  $n_{neurons}$,
		followed with another layer with $N$ dimensional output which after activation, discussed below, provides  the $u$ vector 
		\item 	 a linear, tanh activated layer with  $n_{neurons}$,
followed with another layer with one-dimensional output which after activation, discussed below, provides  the $T$ value 
	\end{enumerate}
	 a dense  layer with $N+1$ dimensional output; the final activation function is described below; according to standard conventions, ‘None’ represents the batch size, which during prediction depends on the user's request.
	\item 'batch size` $n_{batch}= 1024 \cdot 8$; we have also tested with 256 and 1024, as one can imagine, more the better; the value $256$  seems to be the minimum required to obtain qualitatively coherent solutions.
	\item the computation of the integrals in the loss function \eqref{eq:fonctionlosspinn1}-\eqref{eq:fonctionlosspinn2}
	is done by  Monte Carlo sampling.
	For the PDE residuals
	 we used uniform sampling in $x$ and exponential truncated sampling (see \cite{pinn_turinici_icpr2025,TURINICI_pinn_2026})  simulated as: 
	\begin{equation}- 	\frac{1}{\lambda} \log\left[  
		1-U\cdot (1-e^{-\lambda T})
		\right],
		\ \ U= \text{ uniform law on } [0,1].
	\end{equation}
where $\lambda > 0$ is the rate parameter of the truncated exponential distribution (value in Table~\ref{tab:valeursMarshak}); the choice $\lambda = 1/T$ concentrates sampling near $t = 0$, where the solution varies most rapidly. \\ For other losses (the boundary condition loss, initial condition, ...) we used uniform sampling over relevant domains. 
See Table~\ref{tab:valeursMarshak} for various parameter values.
\end{enumerate}

Let us discuss the final activation. 

\begin{enumerate}
	\item 
	{\bf Tanh activation:} the canonical choice of the 'tanh' activation is invalid here because 
	'tanh' cannot represent values outside $[-1,1]$ which are required by 
	the boundary data for $\Ttil$;
	\item 
	{\bf No activation (also called 'linear' activation)}
	This activation has the drawback of producing values that can be negative, which do not comply with the physical interpretation of $\util$ and $\Ttil$. Moreover, dividing by $\Ttil^3$ can become very unstable if it crosses zero (or is around it). Therefore, during convergence it is necessary to use activations that guarantee positivity. This is not surprising, since we are dealing with intensities and temperatures. See Figure~\ref{fig:profile_ablation} for numerical results.
	\item {\bf ReLU family}
	To address the positivity constraint, one option would be to use 
	the famous ReLU function $\text{ReLU}(x) = x_+$ except that 
	its second derivative vanishes almost everywhere so one cannot obtain meaningful information for PDE loss. As an alternative, proposals in the literature include its powers: 
	$\text{ReLU2}(x) =	(\text{ReLU})^2(x)$,  
	$\text{ReLU3}(x) =	(\text{ReLU})^3(x)$.
	Some possible variations include 
	'SafeSoftplus', 
	\begin{align}
	\text{SafeSoftplus}(x)= \frac{1}{\beta_{SSP}} \log\big(1 + \exp(\beta_{SSP} x)\big) + \epsilon_{SSP}  	\end{align}
parameterized by 
	$\epsilon_{SSP} > 0$ and $\beta_{SSP} > 0$. 
	Nevertheless, the results for these functions appeared less satisfactory compared to 
	the next choice.
	\item {\bf ScaledSigmoid:}
	the previous choices do not prevent $\util$ or $\Ttil$ from taking very small values, which causes instabilities; similarly, very large values slow down learning. We therefore looked for activations with output within a prescribed range. In particular we noted that a renormalized sigmoid has good practical properties:
	\begin{align}
		& \text{ScaledSigmoid}(x)=v_{min} + (v_{max}-v_{min})\cdot \frac{1}{1+e^{-x}}.
	\end{align}
	This activation takes values in $]v_{min}, v_{max}[$\footnote{We recover the classical sigmoid function for $v_{min}=0$, $v_{max}=1$.}. Note that, in principle, one or both parameters $v_{min}$ and $v_{max}$ can be learned by the algorithm. In practice, we know that our system satisfies the maximum principle and thus $\util \in [\util_{init},\util_{leftbcu}]^N$  and $\Ttil \in [\Ttil_{init},\Ttil_{leftbct}]$, so we use this a~priori information. The lower bound $v_{min}$ is kept fixed in all runs: some terms of the system involve $1/\Ttil^3$, which is an illegal operation at $\Ttil=0$, so for numerical stability reasons we need to assume $\Ttil$ larger than some positive constant anyway, and training $v_{min}$ over a large range is not useful. Section~\ref{sec:sensitivity} reports a run in which $v_{max}$ is learnable and shows that the results are stable with respect to this choice. See Table~\ref{tab:valeursMarshak} for initial values.
\end{enumerate}

\subsection{Loss function (I): log MSE loss}

As discussed before, one difficulty for this problem comes from the large parameter range, with initial values of order $10^{-12}$ and boundary values of order $10^0$. In this case just fitting the boundary and initial data is very difficult. In practice a quadratic loss will only fit 
$10^0$ values and will have a huge relative order for the  $10^{-12}$ regime.

Our original proposal comes from the idea that we need to have training information on a logarithmic scale, so we looked for a loss function with derivative involving the logarithm of the relative error of the current value with respect to the target.  
To this end we introduce a MSE logarithmic loss. It is defined for positive vectors $\mathbf{x}$ and $\mathbf{y}$ as:
\begin{align}
	\dloss(\mathbf{x} \,\|\, \mathbf{y})
	= \sum_i \left[ \log  (|x_i|+\epsilon_L) - \log(|y_i| + \epsilon_L) 	\right]^2,
	\label{eq:poisson_div_vect}
\end{align}
with $\epsilon_L$ a very small constant (see Table \ref{tab:valeursMarshak}) that ensures the logarithm has finite value.
The following loss function formulation is used for both initial and boundary conditions instead of definitions in 
\eqref{eq:loss_ic_bc}:
\begin{align}
	&\Lcal_{ic}^{\dloss}(u_\theta(0,x) , \xi(x))= \int_\Dcal 	\dloss(u_\theta(0,x), \xi(x) ) dx \\
	&	\Lcal_{bc}^{\dloss}(u_\theta(t,x) , u_{bc}(t,x))=\int_{[0,T] \times\partial\Dcal} 	\dloss(u_\theta(t,x), u_{bc}(t,x) ) dxdt,
	\label{eq:loss_ic_bc_pdiv}
\end{align}

\subsection{Loss function (II): Conservation laws}

Note that the system \eqref{eq:tilde_eq_uT1}-\eqref{eq:tilde_eq_uT_last} admits a conservation law that can be found in the following way; first take the average over $\omega$ of $ \util(t,x,\omega)$ then sum up all equations; the terms 
$\frac{\csigmatil}{\Ttil^3(t,x)} \langle\util\rangle(t,x)$ and $\csbtil \Ttil(t,x)$ will cancel out and we obtain:
\begin{align}
	& \frac{1}{c} \partial_t \langle\util\rangle(t,x)+ 
	\langle\omega \nabla_x \util\rangle(t,x) + 	\cvtil \partial_t \Ttil(t,x) =0.
\label{eq:conservation_law0}
\end{align}
Now integrate over $[x_{min},x_{max}]\times [0,t]$ to obtain:
\begin{align}
	&
	 \frac{1}{c} \int_{x_{min}}^{x_{max}} \langle\util\rangle(t,x) - \langle\util\rangle(0,x) dx +
\int_0^t	 \langle\omega  \util\rangle(\tau,x_{max})- \langle\omega  \util\rangle(\tau,x_{min}) d \tau 
\nonumber \\ &
+	\cvtil  \int_{x_{min}}^{x_{max}} \Ttil(t,x) - \Ttil(0,x) dx=0.
	\label{eq:conservation_law1}
\end{align}

If we consider now that the value $x_{max}$ has been chosen large enough such that on the right border there is still equilibrium i.e. $\util(t,x_{max})=\util_{init} $ and denote 
$e_\omega$ the average over the entrant directions
\begin{align}
	e_\omega= \frac{1}{N} \sum_k \omega_k \cdot 1_{\omega_k \ge0}
\end{align}
we obtain
\begin{align}
 \forall t\le T :~	Cons(\util,\Ttil,t)= 0,
\end{align}
where 
\begin{align}
&	Cons(\util,\Ttil,t):=
\frac{1}{c} \int_{x_{min}}^{x_{max}} (\langle\util\rangle(t,x) - \util_{init} )dx
+	\cvtil  \int_{x_{min}}^{x_{max}} (\Ttil(t,x) - \Ttil_{init}) dx
\nonumber \\ &
+ e_\omega \cdot t \cdot (\util_{init}-\util_{leftbcu})
+ \frac{1}{N} \sum_{\omega \in \Omega_N, \omega <0} \int_0^t \omega\cdot (\util_{init}-\util(\tau,x_{min},\omega))  d\tau.
\label{eq:conservation_law}
\end{align}

\noindent
The conservation laws convey an important part of the system's physics. As such we included it in the $\mathcal{L}_\text{physics}$ loss part (see Equation \eqref{eq:fonctionlosspinn2}). In practice we take $n_{cons}$ time instants $t_1$, ..., $t_{n_{cons}}$ equally distributed between $T/100$ and $T$ and 
add to $\mathcal{L}_\text{physics}$ the average 
of all square conservation law errors $Cons(\util,\Ttil,t_k)^2$. The integrals in 
\eqref{eq:conservation_law}
 are computed through a Monte Carlo sampling (each integral has $n_{MonteCarlo}$ equally spaced samples) and the overall result was multiplied with a coefficient $c_{cons}$ as the other terms in 
 \eqref{eq:fonctionlosspinn2} (value in Table~\ref{tab:valeursMarshak}).

\section{Numerical results} \label{sec:results}

\begin{figure}
\begin{center}
\includegraphics[width=.6\linewidth]{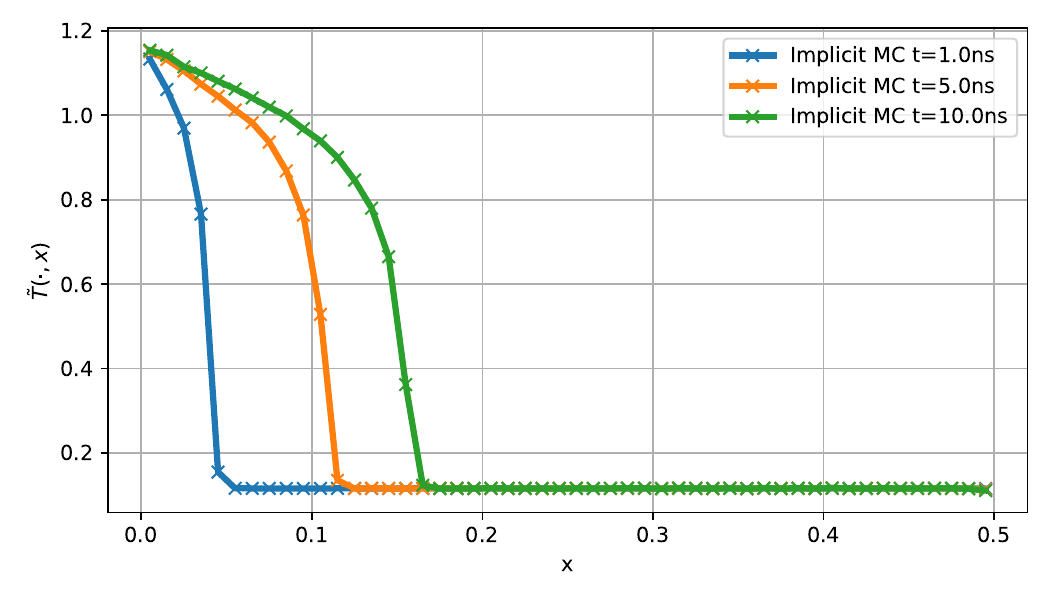}
\end{center}
	\caption{Reference solution for the problem \eqref{eq:tilde_eq_uT1}-\eqref{eq:tilde_eq_uT_last} 
	obtained by the Implicit Monte Carlo method \cite{osti_1255843,FLECK1971313}. 
	For this method the $x$ grid is spaced by $\Delta x = 0.05$ cm and the time step is $\Delta t = 2 \times 10^{-1}$ ns.	
	We see that the Marshak wave reaches $x \sim 0.03cm$ by $1ns$, $\sim 0.08cm$ by $5ns$ and $\sim 0.12-0.14cm$ by $10ns$.
	}
	\label{fig:reference_miniapp}
\end{figure}

The main output of the Marshak wave computation is the temperature evolution $\Ttil$; it is known,
see~\cite{	marshak1958effect,mcclarren2008effects,LAGUZET_Turinici_24_quantization,marshak22_imex,marshak2B_2015,marshak2B_2022}, that, in the $t,x$ space, the temperature dependence can be divided in several regions: 
\begin{itemize}
\item 
a 'cold' region where the previous equilibrium is still in place, i.e. 
$\util \sim u_{init}$ and $\Ttil(t,x) \sim \tinit$; 
\item a 'hot' region where the new equilibrium is observed i.e.
$\util \sim u_{leftbcu}$ and $\Ttil(t,x) \sim \Ttil_{leftbct}$;
\item a transitory region between the two, called the 'wave front'.
\end{itemize}
The quality of the simulation is given by the presence of these three regions and, if possible, by a good estimation of the propagation speed of the 'wave front'.
Such a dynamics is seen in Figure~\ref{fig:reference_miniapp}. 
We will plot the profiles 
$\Ttil(t_\ell,x)$ with $t_\ell\in \mathcal{T}$ taken as representative time instants, see Table~\ref{tab:valeursMarshak} for values.
The result in Figure~\ref{fig:reference_miniapp} will be considered to be the reference solution.

\subsection{Nominal results}

\begin{figure}
	\includegraphics[width=.55\linewidth]{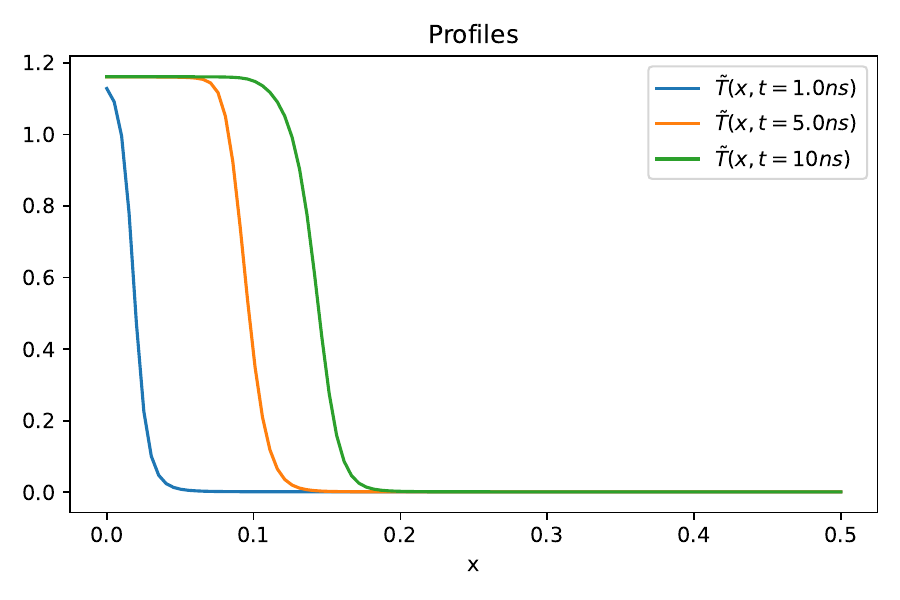} 
	\includegraphics[width=.3\linewidth]{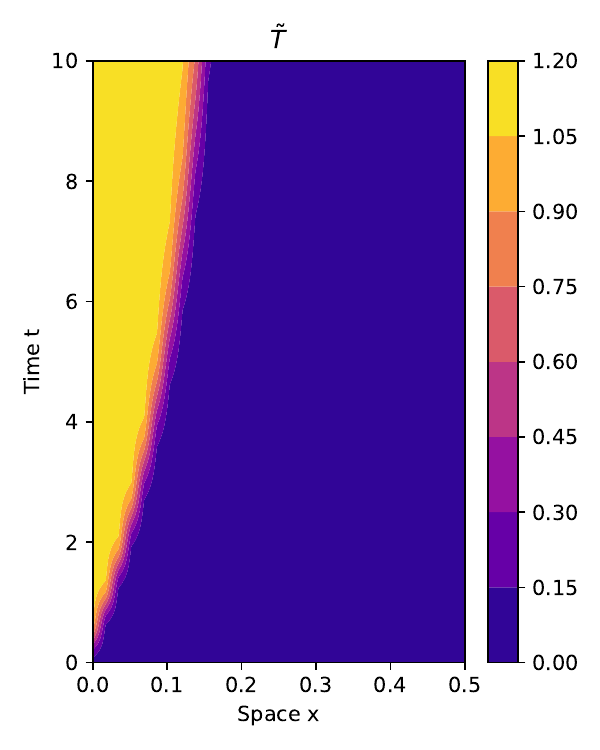}
	\caption{Execution result after $n_{iter}=10k$ iterations using the Adam optimizer (learning rate and other parameters are PyTorch defaults).
{\bf Left:} the Marshak profiles at different time instances; the front position is quantified against the IMC reference in Table~\ref{tab:front_nominal}.
{\bf Right:} the complete $\Ttil$ evolution, time is the ordinate. We note that the $t,x$ space is clearly separated in three regions: 'hot' (yellow), 'wave' (red) and 'cold' (blue).}
	\label{fig:front_Ttilde}
\end{figure}

\begin{figure}
	\includegraphics[width=.24\linewidth]{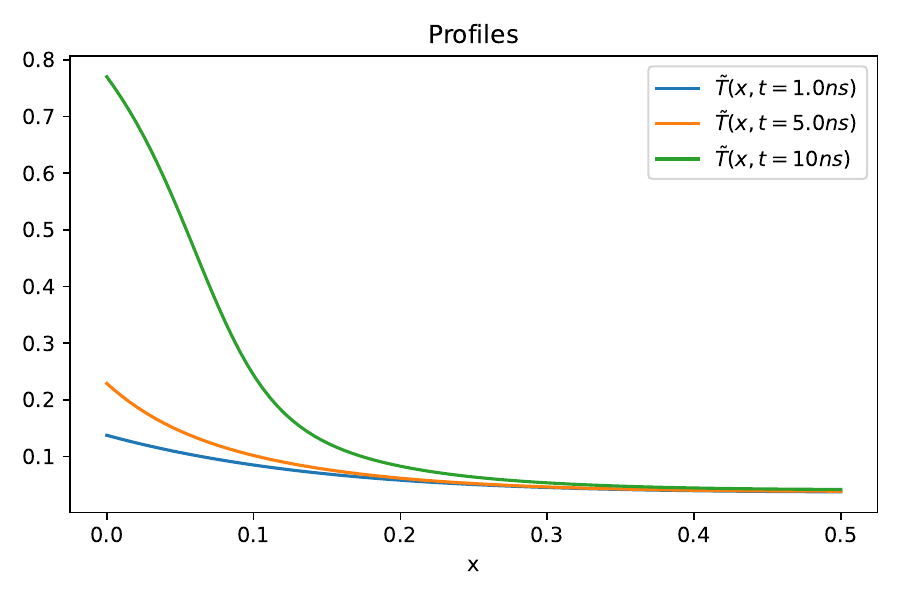} 
	\includegraphics[width=.24\linewidth]{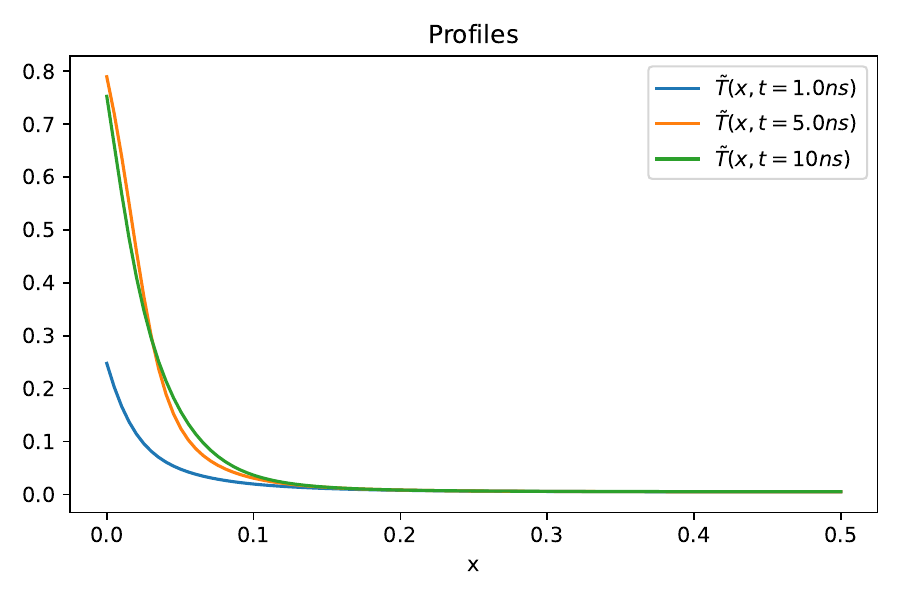} 
	\includegraphics[width=.24\linewidth]{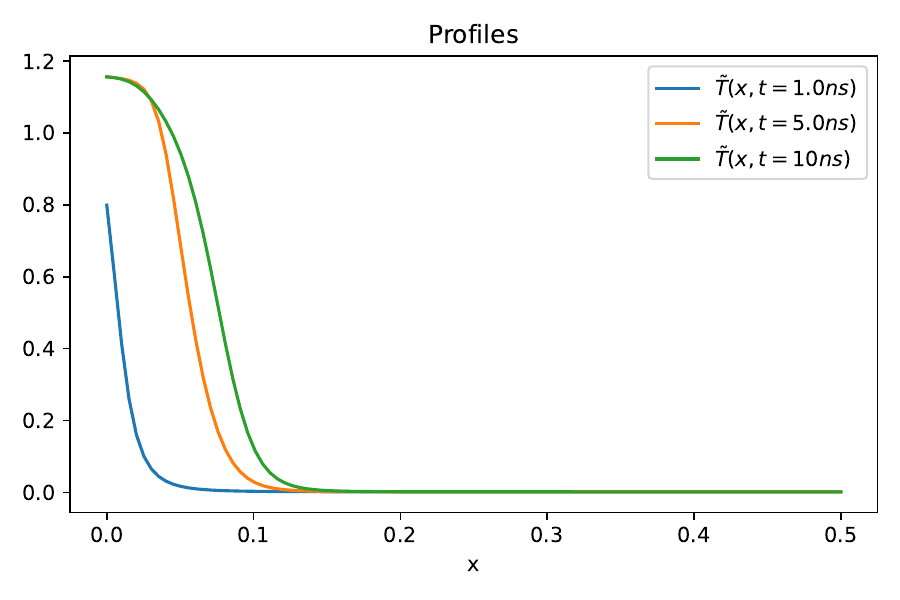} 
	\includegraphics[width=.24\linewidth]{f_g9999profiles_nominal.pdf} 
	\caption{
		Convergence evolution for the result in 
Figure~\ref{fig:front_Ttilde}; 
we plot the profiles for 
$3k$, $5k$, $8k$ 
and ultimately 
$10k$ 
iterations.}
\label{fig:front_conv}
\end{figure}

We present in Figure~\ref{fig:front_Ttilde} the numerical results for our procedure. 
We observe that the results are completely coherent with the reference in Figure~\ref{fig:reference_miniapp}; moreover the Marshak wave and its regions are perfectly clear on the figure. We also plot in Figure 
	\ref{fig:front_conv}
some of the previous iterates; we see that some work is done in order to stabilize the estimations of the Marshak wave but once the solution is found it is stable (despite inherent oscillations present in a stochastic search algorithm).

Overall we interpret this result as positive because of the correct qualitative results; the total run time is less than 5 minutes which is competitive with 
alternatives (finite differences or Implicit Monte Carlo method \cite{osti_1255843,FLECK1971313}); moreover the procedure arrives to handle correctly the $11$ orders of magnitude difference between the term $\util$ of order $10^0$ and 
$\frac{\csigmatil}{\Ttil^3(t,x)}$ of order $10^{11}$.

\paragraph{Quantitative front-position error.}
To quantify the agreement with the reference solution, we track the main quantity of interest, the wave-front position. At each time $t$ we evaluate the front position $x_{0.5}(t)$, the half-height crossing of the normalized profile, for both the PINN solution and the IMC reference (on a common grid, $\Delta x = 0.05$~cm, $\Delta t = 0.2$~ns), and we report the absolute error $|x_{0.5}^{\mathrm{pinn}} - x_{0.5}^{\mathrm{ref}}|$ (in cm) as well as the relative error $|x_{0.5}^{\mathrm{pinn}} - x_{0.5}^{\mathrm{ref}}|/|x_{0.5}^{\mathrm{ref}}|$.

For the nominal case (Table~\ref{tab:front_nominal}), the front position is captured with sub-millimeter absolute error at all reported times. At the earliest reported time, $t=1$~ns, during the ignition delay, the front is still close to the entrance boundary and the relative error is substantial ($57.4\%$): this early-time limitation is acknowledged in Section~\ref{sec:limitations}.

\begin{table}[h]
\centering
\footnotesize
\begin{tabular}{@{}lccccc@{}}
\toprule
$t$ & $x_{0.5}^{\mathrm{pinn}}$ & $x_{0.5}^{\mathrm{ref}}$ & abs.\ error & rel.\ error \\
(ns) & (cm) & (cm) & (cm) & \\
\midrule
1 & $1.93\cdot 10^{-2}$ & $4.53\cdot 10^{-2}$ & $2.60\cdot 10^{-2}$ & 57.4\% \\
5 & $1.05\cdot 10^{-1}$ & $1.04\cdot 10^{-1}$ & $1.16\cdot 10^{-3}$ & 1.1\% \\
10 & $1.57\cdot 10^{-1}$ & $1.48\cdot 10^{-1}$ & $8.62\cdot 10^{-3}$ & 5.8\% \\
\bottomrule
\end{tabular}
\caption{\it Front position (nominal case) compared with the IMC reference solution: absolute error (cm) and relative error.}
\label{tab:front_nominal}
\end{table}

\begin{figure}
	\includegraphics[width=.99\linewidth, trim=0pt 0pt 180pt 0.1pt,clip]{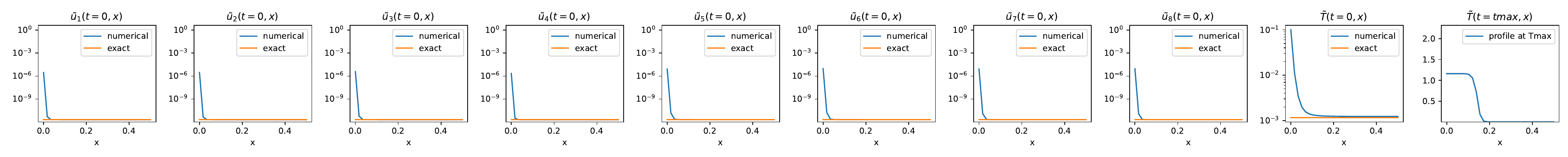}
	
	\includegraphics[width=.99\linewidth, trim=0pt 0pt 0pt 0.1pt,clip]{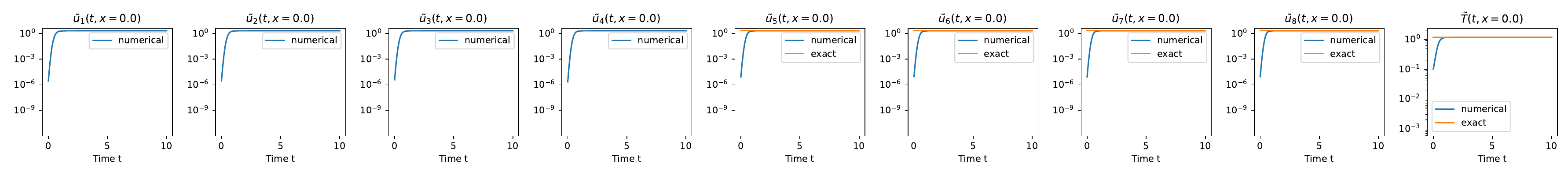}
	\caption{
Initial conditions
and left boundary conditions 
for the result in 
		Figure~\ref{fig:front_Ttilde};  excellent agreement is obtained.
The target values are those with legend 'exact'. 
		Top row of plots: $\util_k(t=0,x)$ $k=1,N$ followed by $\Ttil(t=0,x)$.
Bottom row: $\util_k(t,x=x_{min})$ $k=1, ..., N$ followed by $\Ttil(t,x=x_{min})$. When direction $\omega_k$ is not entering the domain, no boundary condition is to be imposed so no target value is plotted.
	}
	\label{fig:resultatspinn_bc}
\end{figure}

As mentioned before, the initial and boundary conditions are key for obtaining good solutions; to attest on the good agreement with these requirements we plot in Figure~\ref{fig:resultatspinn_bc} the initial and boundary conditions for the solution in Figure~\ref{fig:front_Ttilde}.

\subsection{Ablation studies: linear activation, no PDE, no MSE log loss}

\begin{figure}
	\includegraphics[width=.49\linewidth]{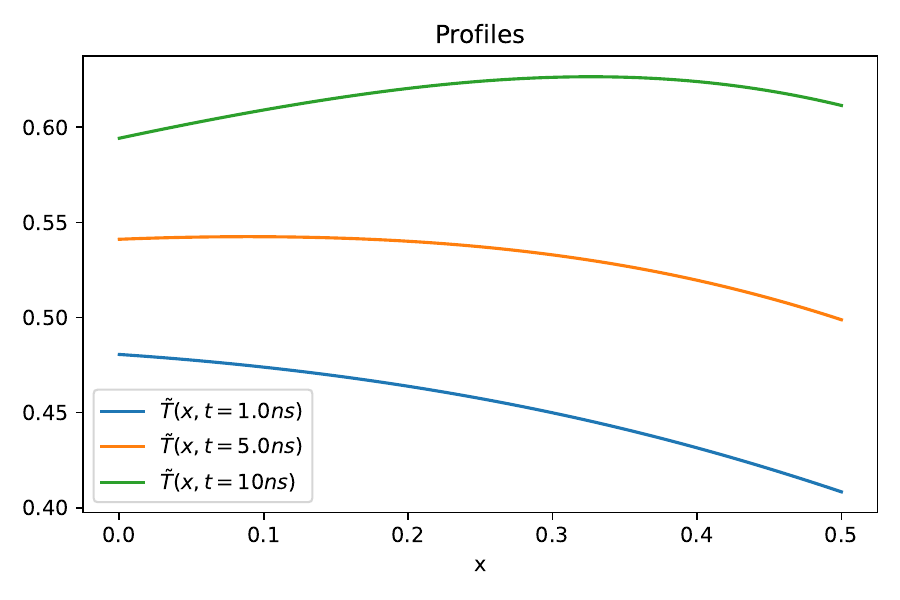} 
	\includegraphics[width=.49\linewidth]{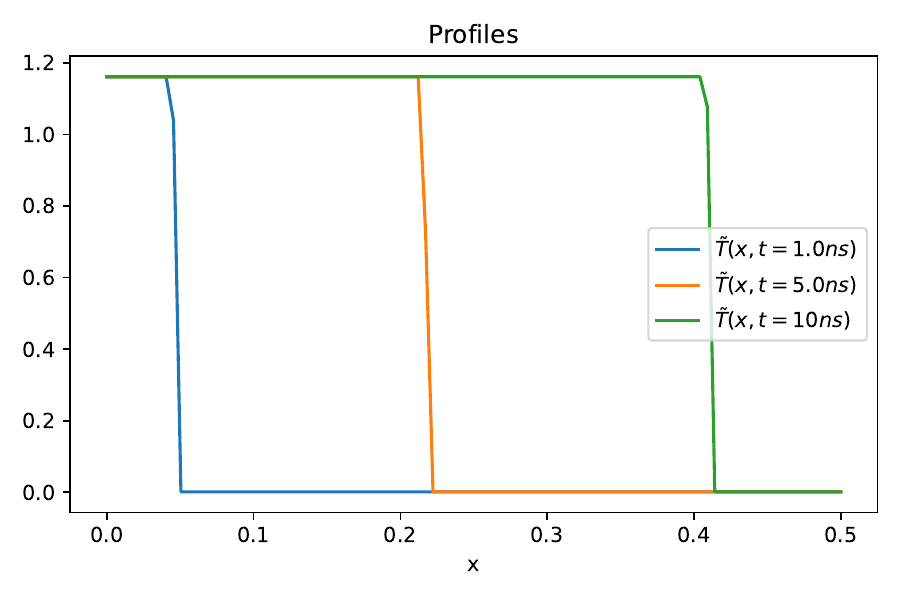} 
	\caption{
Same as Figure~\ref{fig:front_Ttilde} except that 
in each case we take out a single ingredient. Left: we use 'linear' activation; the profiles are clearly not correct;
Right: no PDE (i.e. $c_\phi=0$); the front-position errors are of several tens of percent or more for the ablated variants, see Table~\ref{tab:ablation_metrics}.
}
	\label{fig:profile_ablation}
\end{figure}

\begin{figure}
\includegraphics[width=.99\linewidth, trim=0pt 0pt 180pt 0pt,clip]{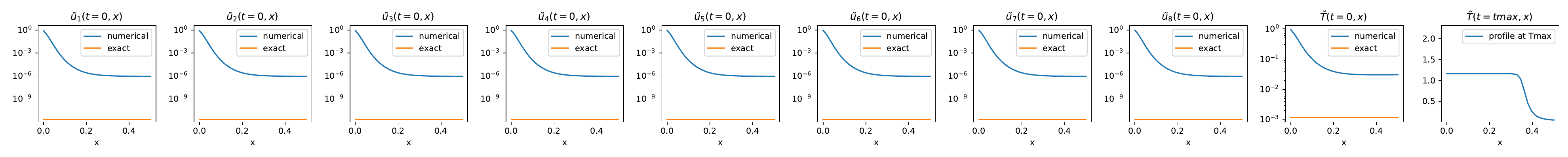}
		\caption{
		Left boundary conditions when replacing the MSE log loss with mean square error. It is seen that, even after $50{,}000$ iterations there is no agreement between the exact and numerical conditions on the left border, with $6$ orders of magnitude mismatch for $\util$ and $1.5$ for $\Ttil$; compare with Figure \ref{fig:resultatspinn_bc}. Obtaining accurate initial conditions is very difficult in this case.
	}
	\label{fig:no_Pdiv}
\end{figure}

We start now a series of comparisons to illustrate the necessity of our main ingredients: the specific activation we use and MSE log loss. The results are presented in Figure
\ref{fig:profile_ablation}. We see that the linear activation is not able to understand the physics of the problem
and 
the solution is not only due to the conservation laws because when eliminating the PDE term the solution becomes unstable. Finally, we see in Figure~\ref{fig:no_Pdiv}
that the lack of MSE log term deteriorates the precision on the initial/ boundary conditions.

The ablation variants are evaluated with the same front-position metric; Table~\ref{tab:ablation_metrics} reports the front position at $t=5$~ns and $t=10$~ns, compared with the IMC reference.

\begin{table}[h]
\centering
\scriptsize
\begin{tabular}{lcccccc}
	\hline
	 & \multicolumn{3}{c}{$t=5$~ns} & \multicolumn{3}{c}{$t=10$~ns} \\
	Configuration & sim            & abs.       & rel.	& sim.           & abs.       & rel.\\
	& $x_{0.5}$ (cm) & error (cm) & error	& $x_{0.5}$ (cm) & error (cm) & error \\
	\hline
	\noalign{\vskip 4pt}

IMC reference & $1.04\cdot 10^{-1}$ & -- & -- & $1.48\cdot 10^{-1}$ & -- & -- \\
Nominal & $1.05\cdot 10^{-1}$ & $1.2\cdot 10^{-3}$ & 1.1\% & $1.57\cdot 10^{-1}$ & $8.6\cdot 10^{-3}$ & 5.8\% \\
Linear activation & $2.28\cdot 10^{-1}$ & $1.24\cdot 10^{-1}$ & 119.1\% & $2.14\cdot 10^{-1}$ & $6.5\cdot 10^{-2}$ & 44.1\% \\
No PDE term & $8.40\cdot 10^{-2}$ & $2.01\cdot 10^{-2}$ & 19.3\% & $8.63\cdot 10^{-2}$ & $6.20\cdot 10^{-2}$ & 41.8\% \\
No log-MSE loss & $2.96\cdot 10^{-1}$ & $1.92\cdot 10^{-1}$ & 184.2\% & $3.35\cdot 10^{-1}$ & $1.87\cdot 10^{-1}$ & 126.1\% \\
\bottomrule
\end{tabular}
\caption{\it Front position (ablation study) at $t=5$~ns and $t=10$~ns, compared with the IMC reference solution: absolute error (cm) and relative error.}
\label{tab:ablation_metrics}
\end{table}

Only the nominal configuration captures the front location with sub-millimeter accuracy; the ablated variants all place the front in the wrong position, with much larger absolute errors.

\subsection{Sensitivity analysis}
\label{sec:sensitivity}

We now study the sensitivity of the front position with respect to several inputs of the framework: the angular discretization order, the material coefficients, the length of the spatial domain, and the treatment of the activation bounds.

\emph{(i) Angular discretization order.} The dependence of the architecture on $N$ is explicit and linear: the output of the $\util$ branch has dimension $N$, the PDE residual is evaluated for the $N$ directions, and the angular average $\langle\util\rangle$ averages $N$ values; the $\Ttil$ branch, the logarithmic MSE loss and the conservation-law loss do not depend on $N$. We re-ran the nominal case with $N=32$, where $15{k}$ iterations were required for convergence: the front dynamics do not change qualitatively, and the front-position errors at $t=5$ and $t=10$~ns remain of the same order as for $N=8$ (Table~\ref{tab:sensitivity}). In practice the solution does not move any more for larger $N$ and $N=8$ is sufficient for this problem.

\emph{(ii) Material coefficients.} We re-ran the nominal case with $\csigmatil$ and $\csbtil$ (the two being consistently rescaled) each decreased by $10\%$, and each increased by $10\%$: the front-position error at $t=5$ and $t=10$~ns stays well below a millimeter in absolute terms (Table~\ref{tab:sensitivity}), i.e., the front is essentially unchanged.

\emph{(iii) Length of the spatial domain.} We re-ran the nominal case with $x_{max}$ varied by $\pm 10\%$ ($0.45$~cm and $0.55$~cm): the front-position error at $t=5$ and $t=10$~ns stays well below a millimeter in absolute terms (Table~\ref{tab:sensitivity}), i.e., the front is essentially unchanged.

\emph{(iv) Learnable activation bounds.} We re-ran the nominal case with the upper bound $v_{max}$ as a trainable parameter in both branches, initialized at the a~priori values $1.001\cdot\util_{leftbcu}$ and $1.001\cdot\Ttil_{leftbct}$. The front-position error is indistinguishable from the nominal case, with sub-millimeter absolute error at both $t=5$ and $t=10$~ns (Table~\ref{tab:sensitivity}): the solution is therefore not an artifact of the tight a~priori bounds. When the physical bounds are known, fixing them to a tight interval is the most efficient option; when the upper bound is unknown, the learnable-bounds variant is the natural mechanism of the framework, and a smooth soft-penalty formulation of the bounds is a natural variant we leave to future work. (The lower bound $v_{min}$ is kept fixed in all runs; see Section~\ref{sec:network_arch} for the reason.)

\begin{table}[h]
\centering
\footnotesize

\begin{tabular}{lcccc}
	\hline
	Configuration & \multicolumn{2}{c}{$t=5$~ns} & \multicolumn{2}{c}{$t=10$~ns} \\
	& abs.\ error (cm) & rel.\ error & abs.\ error (cm) & rel.\ error \\
	\hline \\ [-10pt]
	Nominal & $1.2\cdot 10^{-3}$ & 1.1\% & $8.6\cdot 10^{-3}$ & 5.8\% \\
$N=32$ & $1.3\cdot 10^{-2}$ & 12.2\% & $2.0\cdot 10^{-2}$ & 13.3\% \\
$\csigmatil,\csbtil$ $-10\%$ & $4.5\cdot 10^{-3}$ & 4.1\% & $1.9\cdot 10^{-3}$ & 1.2\% \\
$\csigmatil,\csbtil$ $+10\%$ & $4.5\cdot 10^{-3}$ & 4.6\% & $1.4\cdot 10^{-2}$ & 9.9\% \\
$x_{max}$ $-10\%$ & $5.0\cdot 10^{-3}$ & 4.8\% & $2.6\cdot 10^{-3}$ & 1.7\% \\
$x_{max}$ $+10\%$ & $6.9\cdot 10^{-3}$ & 6.6\% & $1.4\cdot 10^{-2}$ & 9.3\% \\
Learnable $v_{max}$ & $1.4\cdot 10^{-3}$ & 1.3\% & $7.7\cdot 10^{-3}$ & 5.2\% \\
\bottomrule
\end{tabular}
\caption{\it Sensitivity analysis: front position $x_{0.5}(t)$, absolute error (cm) and relative error, at $t=5$~ns and $t=10$~ns. ``Nominal'' is the reference case of Table~\ref{tab:valeursMarshak}; the $N=32$ run required $15{k}$ iterations for convergence; ``$\csigmatil,\csbtil$ $-10\%$/$+10\%$'' means the two coefficients consistently rescaled; ``$x_{max}$ $-10\%$/$+10\%$'' means $x_{max}=0.45/0.55$~cm; ``Learnable $v_{max}$'' uses trainable upper bounds in both branches.}
\label{tab:sensitivity}
\end{table}

\subsection{Time sampling strategy}
\label{sec:sampling}

The PDE residual is sampled with the exponential time density of Section~\ref{sec:methodology} (rate $\lambda=1/T$). We compare it with the uniform time density by re-running the nominal case: with uniform sampling the front-position error is much larger in absolute terms than with exponential sampling, both at $t=5$~ns and at $t=10$~ns (Table~\ref{tab:sampling}). This difference is explained by the structure of the solution: the most rapid temporal variations are concentrated near $t=0$ (ignition and first propagation of the wave), while the uniform density wastes most of its budget on the slowly evolving late regime. We therefore retain the exponential sampling.

\begin{table}[h]
\centering
\footnotesize
\begin{tabular}{lcccc}
	\hline
	Time sampling & \multicolumn{2}{c}{$t=5$~ns} & \multicolumn{2}{c}{$t=10$~ns} \\
	& abs.\ error (cm) & rel.\ error & abs.\ error (cm) & rel.\ error \\
	\hline
	\noalign{\vskip 4pt}
	Exponential (nominal) & $1.2\cdot 10^{-3}$ & 1.1\% & $8.6\cdot 10^{-3}$ & 5.8\% \\
Uniform & $3.1\cdot 10^{-2}$ & 29.4\% & $4.3\cdot 10^{-2}$ & 29.1\% \\
\bottomrule
\end{tabular}
\caption{\it Comparison of the exponential and uniform time sampling strategies on the nominal case (front-position error at $t=5$~ns and $t=10$~ns).}
\label{tab:sampling}
\end{table}

\noindent Adaptive residual-based resampling of the collocation points is a standard technique in the PINN literature for mildly stiff problems. For the present system it proved infeasible in practice: the residual is dominated by the stiff source terms that span $12$ orders of magnitude, and residual-driven resampling concentrates the samples in the most ill-conditioned region of the space--time domain. On a problem where we can barely obtain a stable solution with a fixed, physics-informed sampling strategy, our experiments with adaptive resampling did not converge in a stable way, and we leave this direction to future work.

\subsection{Limitations}
\label{sec:limitations}

Although the front position is captured with sub-millimeter absolute error at all reported times, the relative error remains substantial at early times ($57.4\%$ at $t=1$~ns, Table~\ref{tab:front_nominal}), where the front is still close to the entrance boundary. The front propagation speed and the front sharpness are likewise less accurate: the errors on these quantities are of the order of dozens of percent, and the predicted front is systematically sharper than the IMC reference. Early-time accuracy, and the accuracy of the front speed and of the front sharpness, are thus the main limitations of the current method and the natural target of future improvements.

\section{Conclusion} \label{sec:conclusion}

In this work we addressed the challenging case of stiff, regime-changing transport equations within the PINN framework, using the Marshak wave equation as a representative benchmark. By combining bounded, positivity-preserving activations, a logarithmic MSE loss to handle extreme scale disparities, and the explicit enforcement of global conservation laws, we obtained stable and physically consistent solutions where standard PINN formulations fail.  The numerical results and ablation studies indicate that these ingredients are not merely incremental refinements but play a central role in capturing the correct dynamics of the coupled hyperbolic system. Starting solely from the governing PDEs and associated conservation laws, the proposed PINN framework is able to recover key physical phenomena, such as the Marshak wave, without prior knowledge of the solution structure or ad hoc modifications of the network architecture. The results thus establish a proof of concept for the present one-dimensional Marshak wave problem; extension of the framework to other coupled stiff transport systems, and to more general geometries and parameter regimes, is left to future work. The quantitative evaluation shows that the front location is captured with sub-millimeter absolute error, and identifies the substantial relative error at early times ($57.4\%$ at $t=1$~ns), together with the front speed and sharpness, as the main limitations of the method and the natural targets of future improvements.


\begin{thebibliography}{10}
	\providecommand{\url}[1]{#1}
	\csname url@samestyle\endcsname
	\providecommand{\newblock}{\relax}
	\providecommand{\bibinfo}[2]{#2}
	\providecommand{\BIBentrySTDinterwordspacing}{\spaceskip=0pt\relax}
	\providecommand{\BIBentryALTinterwordstretchfactor}{4}
	\providecommand{\BIBentryALTinterwordspacing}{\spaceskip=\fontdimen2\font plus
		\BIBentryALTinterwordstretchfactor\fontdimen3\font minus
		\fontdimen4\font\relax}
	\providecommand{\BIBforeignlanguage}[2]{{%
			\expandafter\ifx\csname l@#1\endcsname\relax
			\typeout{** WARNING: IEEEtran.bst: No hyphenation pattern has been}%
			\typeout{** loaded for the language `#1'. Using the pattern for}%
			\typeout{** the default language instead.}%
			\else
			\language=\csname l@#1\endcsname
			\fi
			#2}}
	\providecommand{\BIBdecl}{\relax}
	\BIBdecl
	
	\bibitem{raissi2019pinns}
	M.~Raissi, P.~Perdikaris, and G.~E. Karniadakis, ``Physics-informed neural
	networks: A deep learning framework for solving forward and inverse problems
	involving partial differential equations,'' \emph{Journal of Computational
		Physics}, vol. 378, pp. 686--707, 2019.
	
	\bibitem{karniadakis2021physics}
	G.~E. Karniadakis, I.~G. Kevrekidis, L.~Lu, P.~Perdikaris, S.~Wang, and
	L.~Yang, ``Physics-informed machine learning,'' \emph{Nature Reviews
		Physics}, vol.~3, no.~6, pp. 422--440, 2021.
	
	\bibitem{sun2020surrogate}
	L.~Sun, H.~Gao, S.~Pan, and J.~Wang, ``Surrogate modeling for fluid flows based
	on physics-constrained deep learning without simulation data,''
	\emph{Computer Methods in Applied Mechanics and Engineering}, vol. 361, p.
	112732, 2020.
	
	\bibitem{jagtap2020adaptive}
	A.~D. Jagtap, K.~Kawaguchi, and G.~E. Karniadakis, ``Adaptive activation
	functions accelerate convergence in deep and physics-informed neural
	networks,'' \emph{Journal of Computational Physics}, vol. 404, p. 109136,
	2020.
	
	\bibitem{raissi2018hidden}
	M.~Raissi, P.~Perdikaris, and G.~E. Karniadakis, ``Hidden physics models:
	Machine learning of nonlinear partial differential equations,'' \emph{Journal
		of Computational Physics}, vol. 357, pp. 125--141, 2018.
	
	\bibitem{yang2021bpinn}
	L.~Yang, X.~Meng, and G.~E. Karniadakis, ``B-pinns: Bayesian physics-informed
	neural networks for forward and inverse pde problems with noisy data,''
	\emph{Journal of Computational Physics}, vol. 425, p. 109913, 2021.
	
	\bibitem{inversepinn_berardi25}
	\BIBentryALTinterwordspacing
	M.~Berardi, F.~V. Difonzo, and M.~Icardi, ``Inverse physics-informed neural
	networks for transport models in porous materials,'' \emph{Computer Methods
		in Applied Mechanics and Engineering}, vol. 435, p. 117628, 2025. [Online].
	Available:
	\url{https://www.sciencedirect.com/science/article/pii/S004578252400882X}
	\BIBentrySTDinterwordspacing
	
	\bibitem{wang2023when}
	S.~Wang, Y.~Teng, and P.~Perdikaris, ``When and why pinns fail to train: A
	neural tangent kernel perspective,'' \emph{Journal of Computational Physics},
	vol. 449, p. 110768, 2022.
	
	\bibitem{baty2023solvingstiffordinarydifferential}
	\BIBentryALTinterwordspacing
	H.~Baty, ``Solving stiff ordinary differential equations using physics informed
	neural networks (pinns): simple recipes to improve training of
	vanilla-pinns,'' 2023. [Online]. Available:
	\url{https://arxiv.org/abs/2304.08289}
	\BIBentrySTDinterwordspacing
	
	\bibitem{bento2026solvingstiffdarkmatter}
	\BIBentryALTinterwordspacing
	M.~P. Bento, H.~B. Câmara, J.~R. Rocha, and J.~F. Seabra, ``Solving stiff dark
	matter equations via jacobian normalization with physics-informed neural
	networks,'' 2026. [Online]. Available: \url{https://arxiv.org/abs/2602.21988}
	\BIBentrySTDinterwordspacing
	
	\bibitem{hofmann2026mitigatingnumericalstiffnessleastsquares}
	\BIBentryALTinterwordspacing
	P.-A. Hofmann and M.~Hecht, ``Mitigating numerical stiffness in least-squares
	formulations of elliptic pdes for physics-informed neural networks,'' 2026.
	[Online]. Available: \url{https://arxiv.org/abs/2607.02726}
	\BIBentrySTDinterwordspacing
	
	\bibitem{jesser2024numericalrobustnesspinnsmultiscale}
	\BIBentryALTinterwordspacing
	A.~Jesser, K.~Krycki, R.~G. McClarren, and M.~Frank, ``Numerical robustness of
	pinns for multiscale transport equations,'' 2024. [Online]. Available:
	\url{https://arxiv.org/abs/2412.14683}
	\BIBentrySTDinterwordspacing
	
	\bibitem{lu2021deeponet}
	L.~Lu, P.~Jin, G.~Pang, Z.~Zhang, and G.~E. Karniadakis, ``Learning nonlinear
	operators via deeponet based on the universal approximation theorem of
	operators,'' \emph{Nature Machine Intelligence}, vol.~3, no.~3, pp. 218--229,
	2021.
	
	\bibitem{li2021fourier}
	Z.~Li, N.~Kovachki, K.~Azizzadenesheli, B.~Liu, K.~Bhattacharya, A.~Stuart, and
	A.~Anandkumar, ``Fourier neural operator for parametric partial differential
	equations,'' \emph{arXiv preprint arXiv:2010.08895}, 2021.
	
	\bibitem{WANG2022109234}
	\BIBentryALTinterwordspacing
	J.~Wang, X.~Peng, Z.~Chen, B.~Zhou, Y.~Zhou, and N.~Zhou, ``Surrogate modeling
	for neutron diffusion problems based on conservative physics-informed neural
	networks with boundary conditions enforcement,'' \emph{Annals of Nuclear
		Energy}, vol. 176, p. 109234, 2022. [Online]. Available:
	\url{https://www.sciencedirect.com/science/article/pii/S0306454922002699}
	\BIBentrySTDinterwordspacing
	
	\bibitem{liang_continuous_2024}
	\BIBentryALTinterwordspacing
	H.~Liang, Z.~Song, C.~Zhao, and X.~Bian, ``Continuous and discontinuous
	compressible flows in a converging–diverging channel solved by
	physics-informed neural networks without exogenous data,'' \emph{Scientific
		Reports}, vol.~14, no.~1, p. 3822, Feb. 2024. [Online]. Available:
	\url{https://doi.org/10.1038/s41598-024-53680-2}
	\BIBentrySTDinterwordspacing
	
	\bibitem{pinn_gradient_pathologies21}
	\BIBentryALTinterwordspacing
	S.~Wang, Y.~Teng, and P.~Perdikaris, ``{Understanding and Mitigating Gradient
		Flow Pathologies in Physics-Informed Neural Networks},'' \emph{SIAM Journal
		on Scientific Computing}, vol.~43, no.~5, pp. A3055--A3081, 2021. [Online].
	Available: \url{https://doi.org/10.1137/20M1318043}
	\BIBentrySTDinterwordspacing
	
	\bibitem{hard_constraintsPINN21}
	\BIBentryALTinterwordspacing
	L.~Lu, R.~Pestourie, W.~Yao, Z.~Wang, F.~Verdugo, and S.~G. Johnson,
	``Physics-informed neural networks with hard constraints for inverse
	design,'' \emph{SIAM Journal on Scientific Computing}, vol.~43, no.~6, pp.
	B1105--B1132, 2021. [Online]. Available:
	\url{https://doi.org/10.1137/21M1397908}
	\BIBentrySTDinterwordspacing
	
	\bibitem{toscano2024pinnspikansrecentadvances}
	\BIBentryALTinterwordspacing
	J.~D. Toscano, V.~Oommen, A.~J. Varghese, Z.~Zou, N.~A. Daryakenari, C.~Wu, and
	G.~E. Karniadakis, ``From {PINNs} to {PIKANs}: Recent advances in
	physics-informed machine learning,'' 2024. [Online]. Available:
	\url{https://arxiv.org/abs/2410.13228}
	\BIBentrySTDinterwordspacing
	
	\bibitem{marshak1958effect}
	R.~E. Marshak, ``Effect of radiation on shock wave behavior,'' \emph{Physics of
		Fluids}, vol.~1, no.~1, pp. 24--29, 1958.
	
	\bibitem{mcclarren2008effects}
	R.~G. McClarren and R.~B. Lowrie, ``The effects of slope limiting on
	asymptotic-preserving numerical methods for hyperbolic conservation laws,''
	\emph{Journal of Computational Physics}, vol. 227, no.~23, pp. 9711--9726,
	2008.
	
	\bibitem{LAGUZET_Turinici_24_quantization}
	\BIBentryALTinterwordspacing
	L.~Laguzet and G.~Turinici, ``{The Quantization Monte Carlo method for solving
		radiative transport equations},'' \emph{Journal of Quantitative Spectroscopy
		and Radiative Transfer}, vol. 329, p. 109178, 2024. [Online]. Available:
	\url{https://www.sciencedirect.com/science/article/pii/S0022407324002851}
	\BIBentrySTDinterwordspacing
	
	\bibitem{marshak22_imex}
	\BIBentryALTinterwordspacing
	J.~Fu, W.~Li, P.~Song, and Y.~Wang, ``An {Asymptotic}-{Preserving} {IMEX}
	{Method} for {Nonlinear} {Radiative} {Transfer} {Equation},'' \emph{Journal
		of Scientific Computing}, vol.~92, no.~1, p.~27, Jun. 2022. [Online].
	Available: \url{https://doi.org/10.1007/s10915-022-01870-3}
	\BIBentrySTDinterwordspacing
	
	\bibitem{marshak2B_2015}
	\BIBentryALTinterwordspacing
	W.~Sun, S.~Jiang, and K.~Xu, ``An asymptotic preserving unified gas kinetic
	scheme for gray radiative transfer equations,'' \emph{Journal of
		Computational Physics}, vol. 285, pp. 265--279, 2015. [Online]. Available:
	\url{https://www.sciencedirect.com/science/article/pii/S0021999115000121}
	\BIBentrySTDinterwordspacing
	
	\bibitem{marshak2B_2022}
	\BIBentryALTinterwordspacing
	T.~Xiong, W.~Sun, Y.~Shi, and P.~Song, ``High order asymptotic preserving
	discontinuous galerkin methods for gray radiative transfer equations,''
	\emph{Journal of Computational Physics}, vol. 463, p. 111308, 2022. [Online].
	Available:
	\url{https://www.sciencedirect.com/science/article/pii/S0021999122003709}
	\BIBentrySTDinterwordspacing
	
	\bibitem{xingMonteCarlo25}
	\BIBentryALTinterwordspacing
	D.~Xing, ``Research on the monte carlo method for view factor calculation based
	on triangular meshes,'' \emph{Thermal Science and Engineering Progress},
	vol.~65, p. 103907, 2025. [Online]. Available:
	\url{https://www.sciencedirect.com/science/article/pii/S2451904925006985}
	\BIBentrySTDinterwordspacing
	
	\bibitem{carlson_solution_1958}
	B.~Carlson and G.~Bell, ``Solution of the transport equation by the {Sn}
	method,'' Los Alamos Scientific Lab., N. Mex., Tech. Rep., 1958.
	
	\bibitem{pinn_turinici_icpr2025}
	G.~Turinici, ``{Optimal Time Sampling in Physics-Informed Neural Networks},''
	in \emph{Pattern Recognition}, A.~Antonacopoulos, S.~Chaudhuri, R.~Chellappa,
	C.-L. Liu, S.~Bhattacharya, and U.~Pal, Eds.\hskip 1em plus 0.5em minus
	0.4em\relax Cham: Springer Nature Switzerland, 2025, pp. 218--233.
	
	\bibitem{TURINICI_pinn_2026}
	\BIBentryALTinterwordspacing
	------, ``Regime-aware time weighting for physics-informed neural networks,''
	\emph{Journal of Computational and Applied Mathematics}, vol. 473, p. 116858,
	2026. [Online]. Available:
	\url{https://www.sciencedirect.com/science/article/pii/S0377042725003723}
	\BIBentrySTDinterwordspacing
	
	\bibitem{osti_1255843}
	\BIBentryALTinterwordspacing
	A.~B. Wollaber, ``{Four decades of implicit Monte Carlo},'' \emph{Journal of
		Computational and Theoretical Transport}, 02 2016. [Online]. Available:
	\url{https://www.osti.gov/biblio/1255843}
	\BIBentrySTDinterwordspacing
	
	\bibitem{FLECK1971313}
	\BIBentryALTinterwordspacing
	J.~Fleck and J.~Cummings, ``{An implicit Monte Carlo scheme for calculating
		time and frequency dependent nonlinear radiation transport},'' \emph{Journal
		of Computational Physics}, vol.~8, no.~3, pp. 313--342, 1971. [Online].
	Available:
	\url{https://www.sciencedirect.com/science/article/pii/0021999171900155}
	\BIBentrySTDinterwordspacing
	
\end{thebibliography}


\end{document}